\documentclass[manuscript]{aastex}
\usepackage{epsfig}                   %%
\usepackage{multirow}
\usepackage{natbib}
\usepackage{txfonts}                  %%
\shorttitle{Effects of rotation in AGB stars}
\shortauthors{Piersanti L., Cristallo S., Straniero O.}

\begin{document}
\title{The effects of rotation on the s-process nucleosynthesys in Asymptotic Giant Branch stars}

\author{L. Piersanti}
     \affil{INAF-Osservatorio Astronomico di Collurania, via Maggini snc, 64100, Teramo, Italy}
       \email{piersanti@oa-teramo.inaf.it}
\author{S. Cristallo}
     \affil{INAF-Osservatorio Astronomico di Collurania, via Maggini snc, 64100, Teramo, Italy}
\author{O.~Straniero}
     \affil{INAF-Osservatorio Astronomico di Collurania, via Maggini snc, 64100, Teramo, Italy}

\date{\today}

\begin{abstract}
In this paper we analyze the effects induced by rotation on low
mass Asymptotic Giant Branch stars. We compute two sets of models,
M=2.0 M$_\odot$ at [Fe/H]=0 and M=1.5 M$_\odot$ at [Fe/H]=-1.7,
respectively, by adopting Main Sequence rotation velocities in the
range 0$\div$120 km/s. At high metallicity, we find that the
Goldreich-Schubert-Fricke instability, active at the interface
between the convective envelope and the rapid rotating core,
contaminates the $^{13}$C-pocket (the major neutron source) with
$^{14}$N (the major neutron poison), thus reducing the neutron
flux available for the synthesis of heavy elements. As a
consequence, the yields of heavy-s elements (Ba, La, Nd, Sm) and,
to a less extent, those of light-s elements (Sr, Y, Zr) decrease
with increasing rotation velocities up to 60 km/s. However, for
larger initial rotation velocities, the production of light-s and,
to a less extent, that of heavy-s begins again to increase, due to
mixing induced by meridional circulations. At low metallicity, the
effects of meridional circulations are important even at rather
low rotation velocity. The combined effect of
Goldreich-Schubert-Fricke instability and meridional circulations
determines an increase of light-s and, to a less extent, heavy-s
elements, while lead is strongly reduced. For both metallicities,
the rotation-induced instabilities active during the interpulse
phase reduce the neutrons-to-seeds ratio, so that the
spectroscopic indexes [hs/ls] and [Pb/hs] decrease by increasing
the initial rotation velocity. Our analysis suggests that rotation
could explain the spread in the s-process indexes, as observed in
s-process enriched stars at different metallicities.

\end{abstract}

\maketitle

\section{Introduction}\label{intro}
Rotation modifies the physical structure and the chemical
composition of a star, essentially because of the lifting due to
the centrifugal force and of the mixing induced by  dynamical and
secular instabilities. The lifting effects on AGB stars have been
firstly studied by \citet{dom96}, who showed that it causes a
delay of the second dredge up in intermediate mass stars and, in
turn, allows the formation of CO cores very close to the
Chandrasekhar limit for a non-rotating degenerate stellar
structure. More recently, models of low-mass AGB stars including
the effects of rotation have been discussed by several authors
\citep{la99,he03,siess04}. Due to the relatively small initial
angular momentum of the majority of low mass stars, lifting
effects are generally negligible in single stars with mass lower
than 3 M$_\odot$. Nonetheless, mixing induced by rotation may have
important consequences on the nucleosynthesis. Indeed, low-mass
AGB stars play a fundamental role in the chemical evolution of
their host galaxy, being the main producers of about half of the
elements heavier than iron. They also contribute to the synthesis
of some light elements, like carbon or fluorine. During the late
part of the AGB phase, these stars undergo recursive thermal
pulses, {\it i.e.} thermonuclear runaways of the He-burning layer
(Thermally Pulsing Asymptotic Giants Branch stars, TP-AGB stars).
If the envelope mass exceeds a critical value, the thermal pulse
is followed by a third dredge up episode (TDU), {\it i.e.} a deep
penetration of the convective envelope into the H-exhausted region
(for a review see \citealt{ir83} and \citealt{stra06}). As a
consequence, the products of the internal nucleosynthesis appear
at the surface. In addition, when the convective envelope recedes,
a thin zone characterized by a variable profile of hydrogen is
left on the top of the H-exhausted core. During the time that
elapses between two successive thermal pulses (interpulse period),
this zone contracts and heats up, until  $^{13}$C and $^{14}$N are
produced as a consequence of an incomplete CN cycle. The resulting
$^{13}$C enriched zone is often called  the $^{13}$C pocket. Just
above this layer, a $^{14}$N pocket forms. As firstly shown by
\citet{stra95} (see also \citealt{ga98}), when the temperature
approaches $\sim$90 MK, the $^{13}$C$(\alpha,n)^{16}$O reaction
provides the neutron flux needed to activate an efficient
s-process nucleosynthesis. \citet{ga98} demonstrate that about
$10^{-5}$ M$_\odot$ of $^{13}$C are required to reproduce the
heavy element enhancements commonly observed in evolved AGB and
post-AGB stars (\citealt{lambert95,abia01,abia02,rey04}). As a
matter of fact, the theoretical tools commonly used to model AGB
stars do not provide a precise quantitative description of the H
abundance left by the TDU in the transition zone between the fully
convective envelope and the radiative core. In turn, the extension
of the $^{13}$C pocket is not very well constrained on the base of
first principles. This problem has been largely discussed in the
community of stellar modelers. One of the main open-questions
concerns the physical process(es) that determines such a H
profile. \citet{he97} firstly supposed that the driving process is
the convective overshoot. Then, to model such an overshoot, they
adopted an exponential decay of the diffusion coefficient at the
convective boundaries. \citet{beckibe}, \citet{caste90} and
\citet{mowlavi} noted that the inner border of the convective
envelope becomes dynamically unstable when it penetrates an
He-rich (or a H-exhausted) region, as it happens during a TDU
episode. In our previous works (for details see \citealt{stra06}),
in order to handle such an instability, we followed an approach
similar to the one introduced by \citet{he97}, by assuming an
exponential decay of the average convective velocity at the inner
border of the convective envelope. The calibration of the free
parameter, namely the strength of the exponential decay, has been
discussed in \citet{cri09}. \citet{la99} investigated the
possibility that rotation-induced mixing, rather than convective
instabilities, may generate a variable H profile on the top of the
H-exhausted core. Note that the mixing of protons should occur
within the relatively short period of time elapsing between the
occurrence of the TDU and the restart of the H-burning, when the
$^{13}$C pocket develops. More recent studies have shown that a
Goldreich-Schubert-Fricke (hereinafter GSF) instability actually
arises during this period of time, but the resulting amount of
$^{13}$C is about one order of magnitude smaller than that
required to reproduce the observed overabundances of heavy
elements \citep{he03,siess04}. Although these studies cast doubts
on the possibility that rotation alone could produce a sizeable
$^{13}$C pocket, nonetheless mixing induced by rotation may still
play a significant role in the s-process nucleosynthesis in low
mass AGB stars. Independently on the processes that determine the
variable H pattern after a TDU episode\footnote{For completeness,
we recall that in addition to convection and rotation, other
physical processes have been considered as possible origin of the
$^{13}$C pocket \citep[{\it e.g.}][]{deto03}}, rotation-induced mixing may
spread out the $^{13}$C pocket. In that case, owing to the
contamination of the $^{13}$C-rich zone with a significant amount
of $^{14}$N, most of the neutrons released by the
$^{13}$C$(\alpha,n)^{16}$O will be captured by the $^{14}$N, thus
reducing the s-process production of heavy elements
\citep{siess04}. In addition, since the $^{13}$C spreads
over a larger zone, the average value of the
neutrons-to-seeds ratio is expected to decrease. As a result, the
abundance of light-s, namely Sr, Y, Zr (hereinafter ls), heavy-s,
namely Ba, La, Nd, Sm (hereinafter hs) and lead\footnote{The
definition of ls and hs may vary from author to author. Usually,
these quantities identify elements with a magic isotope, {\it
i.e.} a nucleus having a particularly stable nuclear structure.}
could be strongly modified.

The purpose of this paper is to investigate the effects of the
rotation-induced instabilities on the nucleosynthesis occurring in
low-mass TP-AGB stars. More specifically, we add rotation-induced
mixing to the other mixing already considered in our previous
works, those due to convection in particular. As in our previous
works, the leading process determining the formation of the
$^{13}$C pocket is the dynamical instability affecting the inner
border of the convective envelope during a TDU episode. Then, we
allow rotation to possibly modify the physical conditions, such as
temperature and density, as well as the chemical stratification of
the layers where neutrons are released and the s process takes
place.

\section{Algorithm and input physics}\label{algo}

The large majority of stars belonging
to the Galactic Disk with mass M$< 3$M$_\odot$
show rotation velocities $<80$ km/s, with mean
values varying from author to author (see, {\it e.g.},
\citealt{groot96,demede99,holmberg07,reiners12}). Since more
compact stars are expected at low metallicities, Halo stars in the
same range of mass were, possibly, faster rotators. We have
computed two sets of stellar evolutionary models for two different
chemical compositions. The first set has solar-scaled composition
with Y=0.269 and [Fe/H]=0 (corresponding to Z=1.4$\times
10^{-2}$). The second set has Y=0.245, [Fe/H]=-1.7 and
[$\alpha$/Fe]=0.5 (corresponding to Z=7$\times 10^{-4}$). The
initial masses are M=2.0 M$_\odot$ for the high metallicity set
and M=1.5 M$_\odot$ for the low metallicity one. Each sequence
starts with a contracting pre-Main-Sequence fully-homogeneous
model and we follow its evolution up to the AGB tip. At the Zero
Age Main Sequence, models are assumed to uniformly rotate, with
initial rotation velocities $v^{ini}_{rot}$=0, 10, 30, 60 or 120
km/s for both the metallicity sets. The initial angular velocity
$\omega_0$ and total angular momentum J$_0$ of each model are
listed in Table \ref{tab1}.

All the models have been computed with the evolutionary code
described in \citet{stra06}, modified to account for rotation. The
adopted nuclear network includes about 500 isotopes (from $^{1}$H
to $^{209}$Bi), coupled by more than 1000 reactions. The physical
inputs, such as the Equation of State, the radiative and the
conductive opacity, the strong and the weak reaction rates, are
the same as in \citet{cri09}. The solar composition has been taken
from \citet{lo2003}. Accordingly, the value of mixing length
parameter, namely $\alpha$=2.1, has been derived by computing a
standard solar model as described in \citet{pi07}.

The effects of rotation on the structural equations have been
accounted for according to the procedure described by
\citet{es76}. In particular, for the gravitational plus rotational
potential we adopt the approximation by \citet{kito70}, {\it i.e.}
the radial component of the centrifugal force is averaged over a
sphere \citep[see][]{piers03, piers10,piers13}. In addition, a
corrective term for the radiative temperature gradient is included
in the energy transport equation \citep{es76}. In order to compute
realistic models one has also to include a description of the
angular momentum transport as determined by both convection and
rotation-induced instabilities. In addition, the angular momentum
profile within a structure as a function of time determines
whether or not macroscopic mass motions occur, thus also affecting
the chemical stratification.

The treatment of angular momentum transport is currently one of
the main problems of evolutionary models computed with 1D
hydrostatic codes including rotation. In fact, the efficiency of
rotation-induced instabilities relies on phenomenological theories
so that they have to be regarded as order-of-magnitude estimates.
Such an approach inevitably implies the introduction of free
parameters, which require an appropriate calibration\footnote{Note
that this approach is similar to the one commonly used to
calibrate the efficiency of convective transport, as it is done,
for example, in the case of the mixing length theory.}.
Observations can be used to constraints the adopted efficiency
even if, up to now, direct measurements of the internal angular
momentum are available only for a small sample of low mass Main
Sequence (MS) and Red Giant (RG) stars. The data obtained by
KEPLER for a sample of RG stars with masses around 1.3 M$_\odot$
and metallicity close to solar show that the ratio between the
angular velocity of the He core and of the surface is about 5-10,
while theoretical models including the effects of rotation predict
a definitely larger value (up to 1000; see for instance
\citealt{egge12,marques13}). Moreover, the experimental
determination of the rotation velocity for a sample of DA White
Dwarfs (WDs) provides a value lower than 10 km/s \citep{berger05},
while the extant theoretical models of low mass AGB stars predict
rapidly spinning CO cores. Such an evidence clearly suggest that
in the current modelling of the evolution of rotating stars either
the transport of angular momentum is underestimated by at least
two orders of magnitude \citep{marques13} or some additional
mechanisms responsible for an efficient extraction of angular
momentum from the core are currently neglected. Magnetic torque
could be a possible solution \citep{spruit2002,suijs08}, although
it is not clear if the dynamo process could be active also in the
radiative envelope of MS stars (see the discussion in
\citealt{langer2012}). Nonetheless, the narrow extension of the
solar tacochline suggests the presence in the radiative inner
zones of a fossil magnetic field  \citep{goug98}, which could
brake the solar core by magnetic torque
\citep{spruit98,spruit2002}. However, \citet{depi07} argued that a
solar models including only this mechanism would result in a
rapidly rotating core, in contradiction with helioseismic data.
They concluded ''that the Tayler-Spruit mechanism may be important
for envelope angular momentum transport but that some other
process must be responsible for efficient spin-down of stellar
cores''. Alternatively, gravity waves could extract angular
momentum from the inner zones of a star determining the observed
features \citep[see][and references therein]{zahn97}. Other
constraints on the angular momentum transport and, hence, on the
corresponding mixing efficiency can be put basing on chemical
abundances observed in stars. The solar lithium abundance seems to
indicate that the mixing related to rotation-induced instabilities
has an efficiency smaller than the one for the transport of
angular momentum \citep{pins89}. Moreover, the observed nitrogen
surface enrichment in massive stars \citep{magilla} is considered
an indication that the damping effect of mean molecular weight
($\mu$) gradient on meridional circulations is smaller than the
current estimates \citep{heger00,yoon06,brott11}. Summarizing, the
issue of the actual efficiency of the
angular momentum transport in stars is, so far, an open problem.\\
In this paper we adopt a common procedure based on diffusion
algorithms to describe the angular momentum transport and the
mixing of the chemical species induced by rotation. Following
\citet{pins89} (see also \citealt{heger00}, \citealt{yoon06} and
\citealt{brott11}) we introduce some free parameters to account
for the many uncertainties due to rotation instabilities and to
additional phenomena, such has the magnetic braking, not
explicitly included.

Following \citet{es78}, we adopt a non-linear diffusion equation
to describe the transport of angular momentum:
\begin{equation}
\left({\frac{\partial\omega}{\partial t}}\right)={\frac{1}{i}}{\frac{\partial}{\partial m}}
\left[(4\pi r^2\rho)^2 i D_J\left({\frac{\partial\omega}{\partial m}}\right)\right]
\label{e:angev}
\end{equation}
where $i$ is the specific moment of inertia, $\omega$ the angular
velocity (in rad/s), $\rho$ the density (in g $\cdot$ cm$^{-3}$)
and $r$ and $m$ the radius and mass coordinate, respectively.
$D_J$ is the total diffusion coefficient for angular momentum
transport defined as the sum of the diffusion coefficients related
to convection and both secular and dynamical rotation-induced
instabilities:
\begin{equation}
D_J=D_{conv}+f_\omega\times(D_{ES}+D_{GSF}+D_{SS}+D_{DS}+D_{SH})
\label{eqdj}\end{equation} where $D_{ES},\ D_{GSF},\ D_{SS},\
D_{DS},\ D_{SH}$ are the diffusion coefficients related to
Eddington-Sweet, Goldreich-Schubert-Fricke, Secular Shear,
Dynamical Shear and Solberg-H\"oiland instabilities, respectively.
At the center ($m=0$) and at the surface ($m=M$) reflecting
conditions are assumed. Such a formulation guarantees the
conservation of the total angular momentum and enforces rigid
rotation in the zones where the diffusion timescale is shorter
than the ones for a structural change of the star. The parameter
$f_\omega$ accounts for possible uncertainties affecting the
angular momentum transport.

The diffusion coefficient for convection is computed according to
the mixing length theory \citep{cox68}. The other diffusion
coefficients are given by the product of a scale length ($l$) and
a characteristic circulation velocity $v$. According to
\citet{es78}, the scale length is defined as:
\begin{equation}
l=\min\left[\Delta r,\vert{\frac{\partial \ln v}{\partial
r}}\vert^{-1}\right]
\end{equation}
where $\Delta r$ is the extension of the unstable zone.

Variations of the chemical species, as due to convection and
rotation-induced mixing, is also described by means of a nonlinear
diffusion equation:
\begin{equation}
\left({\frac{\partial X_k}{\partial
t}}\right)={\frac{\partial}{\partial m}} \left[(4\pi r^2\rho)^2
D_C\left({\frac{\partial X_k}{\partial m}}\right)\right]
\label{e:chemev}
\end{equation}
where $X_k$ represent the mass fraction of the $k$ chemical
species. In order to take into account the variations of $\omega$
and $\mu$ into the diffusion coefficients, Eqs. (\ref{e:angev})
and (\ref{e:chemev}) should be coupled.

The total diffusion coefficient in  Eq. (\ref{e:chemev}) is:
\begin{equation}
D_C=D_{conv}+f_\omega\times
f_c\times\left(D_{ES}+D_{GSF}+D_{SS}+D_{DS}+D_{SH}\right)
\label{eqdc}\end{equation} where $f_c$ is intended to reduce the
matter circulation with respect to the angular momentum transport,
and should be $\le 1$. \citet{pins89} obtained $f_c=0.046$ by
fitting the solar lithium surface abundance. The physical
motivation for such a factor was provided by \citet{chabo92}, who
showed that, under the hypothesis of shellular rotation,
horizontal turbulence inhibits vertical mixing whereas angular
momentum transport is not affected. As a consequence, in an
Eddington-Sweet unstable zone meridional circulations arise,
determining an angular momentum transport more efficient than the
corresponding mixing of chemical species. According to
\citet{chabo92}, the ratio of the diffusion efficiency over the
turbulent viscosity is $f_c=1/30$. Such a value has been adopted
by many authors
\citep{heger00,yoon04,heger05,petro05,yoon05,yoon06,suijs08}, even
if a slightly lower $f_c$ has been adopted by \citet{brott11} to
reproduce the surface abundances of a B-stars sample in the Large
Magellanic Cloud obtained by the FLAMES survey ($f_c$=0.028). Note
that, in the previous listed works, as well as in the present one,
such a parameter has been used to reduce the mixing efficiency of
all the rotation-induced instabilities even if, in principle, it
should be applied to meridional circulations only.

As it will be shown in the following sections, the rotation
instabilities affecting the s-process nucleosynthesis during the
AGB evolution of low mass stars are the Eddington-Sweet and the
Goldreich-Schubert-Fricke instabilities. The meridional
circulations velocity of a mass flow can be derived by estimating
the mass flux needed to balance the von Zeipel effect
\citep{kipp74} and in chemically homogeneous regions it is given
by the following relation:
\begin{equation}
v_{ES}={\frac{\nabla_{ad}}{\delta(\nabla_{ad}-\nabla)}}{\frac{\omega^2 r^3 L}{(Gm)^2}}
\left[{\frac{2\varepsilon r^2}{L}}-{\frac{2 r^2}{m}}-   {\frac{3}{4\pi\rho r}}\right]
\label{e:ved}
\end{equation}

\cite{gs67} and independently \cite{fricke68} demonstrated that in
radiative chemically-homogeneous zones differential rotation is
stable when the specific angular momentum is an increasing
function of the radius. By assuming cylindrical symmetry this
corresponds to the following conditions:
\begin{equation}
{\frac{\partial j}{\partial z}}=0;\ \ \ \ {\frac{\partial j}{\partial w}}\ge 0
\end{equation}
where $w$ is the distance from the axis of rotation. Such a
condition is also sufficient when considering axisymmetric
perturbations. \cite{james71} provided an estimate for the
large-scale circulation velocity, that in the equatorial plane is
given by:
\begin{equation}
v_{GSF}= {\frac{2 H_T r}{H_j}}\left({\frac{2\omega}{r{\frac{\partial\omega}{\partial r}}}+1}\right)^{-1} v_{ES}
\label{e:vgsf}
\end{equation}
where $H_T$ and $H_j$ are the temperature and specific angular
momentum scale height, respectively, and $v_{ES}$ is the
Eddington-Sweet circulation velocity. By an inspection to Eqs.
(\ref{e:ved}) and (\ref{e:vgsf}), it comes out that the ES
circulation dominates in zones where the angular velocity gradient
is small, while GSF circulation becomes an important mechanism in
zones with a steep $\mu$ and $\omega$ gradient, as in the case of
radiative regions located close to a convective boundary.

As usual, we model the effect of $\mu$-gradient barriers by
defining an equivalent $\mu$-current which works against the
rotation-induced circulation. The corresponding velocity has been
estimated according to \cite{kipp74}:
\begin{equation}
v_{\mu}=f_\mu{\frac{H_P}{\tau_{th}}}{\frac{\varphi\nabla_{\mu}}{\nabla-\nabla_{ad}}}
\label{eqvmu}\end{equation} where $H_P$ is the pressure scale
height, $\nabla_\mu$ is the mean molecular weight gradient,
$\tau_{th}$ is the thermal relaxation timescale of a bubble having
a different $\mu$ with respect to the surrounding and $\varphi$
is, as usual, $(\partial\ln\rho/\partial\ln\mu)_{P,T}$. According
to the previous discussion, the parameter $f_\mu$ has been
introduced to explore the sensitivity of the obtained results on
the adopted formulation for the $\mu$ currents.\\
In the extant literature different values have been adopted for
such a parameter: 1.00 \citep{pins89}, 0.05 \citep{heger00}, 0.10
\citep{yoon06,brott11}. As noted by \cite{heger00}, according to
the available abundance determinations in massive stars, it is not
possible to define the values of both $f_c$ and $f_\mu$
univocally, as ''{\sl different combinations might result in
similar surface enrichments}'', even if the corresponding angular
momentum profiles could be different \citep[see also the
discussion in][]{chi13}.

In our standard case we adopt
$f_\mu$=$f_c$=$f_\omega$=1. In \S \ref{unce} we discuss the effects
induced by a variation of these parameters on the s-process
nucleosynthesis.

The models computed in the present work does not have a convective
envelope during the MS phase, so that we neglect the magnetic
braking by stellar winds \citep{kawa88}.

Mass loss rate is computed according to the Reimers' formula
\citep{reimers75}, by taking $\eta=0.4$, except for the AGB phase
where we adopt the mass-loss rate described in \citet{stra06}.
Accordingly, we assume that the matter lost ($\Delta M$) carries
away an amount of angular momentum given by:

\begin{equation}
\Delta J=\int^M_{M-\Delta M}j(m)dm
\end{equation}

\noindent where $M$ is the total mass of the star and $j(m)$ is
the specific angular momentum.

As recalled in \S \ref{intro}, during the TDU a sharp
discontinuity in the chemical composition takes place at the
convective-radiative boundary, causing the onset of a dynamical
instability. In order to handle this instability, \cite{stra06}
introduced a transition zone between the fully convective envelope
and the radiative core, where the convective velocity
exponentially drops from about $10^5$ cm/s to zero. The strength
of the exponential decaying velocity has been calibrated by
\cite{cri09}, who derived a best value of 0.1 pressure scale
height. The need for such a transition zone still remains in
rotating models, because the velocities of rotation-induced
instabilities are 3-4 orders of magnitude smaller than those of
convection. In the AGB models here presented, mixing and angular
momentum transport in the convective envelope are treated
according to the time-dependent algorithm described in
\cite{stra06}.

\section{Results}\label{res}

\subsection{Evolution before the TP-AGB phase}\label{preagb}

\begin{figure}
\includegraphics[angle=0,width=\columnwidth]{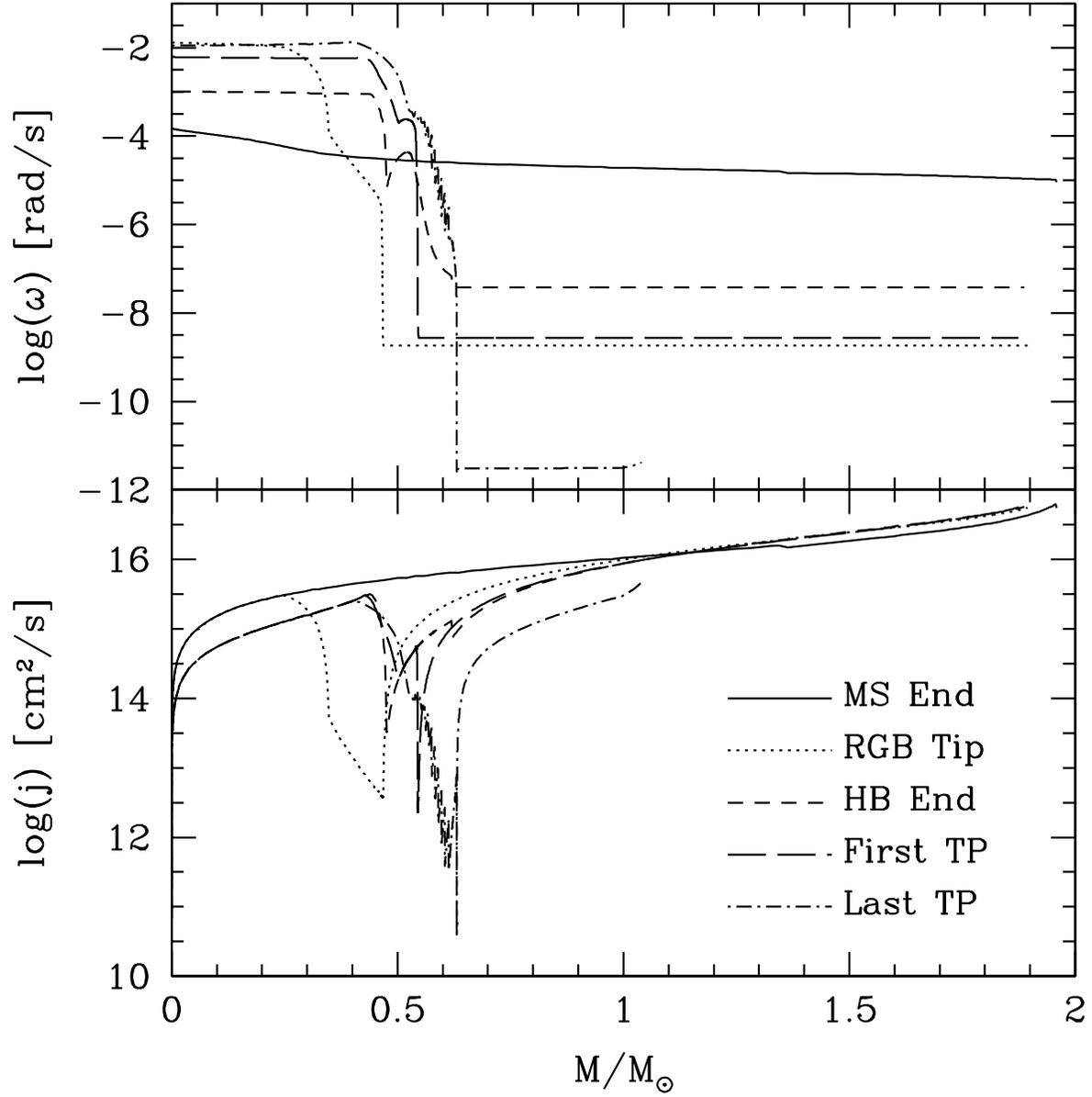}
\caption{Evolution of angular velocity (upper panel) and specific
angular momentum (lower panel) for the model with M=2 M$_\odot$,
[Fe/H]=0 and $v^{ini}_{rot}=30$ km/s. Each curve refers to a
different epoch (see text for details).}\label{fig1}
\end{figure}

The evolution of angular velocity and angular momentum for the
model with M=2 M$_\odot$, [Fe/H]=0 and $v^{ini}_{rot}=30$ km/s is
illustrated in Figure \ref{fig1}. After the exhaustion of the
central H, the core contracts and, in turn, spins up. On the
contrary, the external layers expand (Red Giant Branch, RGB), so
that the angular velocity of the whole convective envelope
substantially decreases. Nevertheless, the overall characteristics
of the computed evolutionary sequences are very marginally
affected by rotation. This is shown in Table \ref{tab1}, where we
report as a function of the mass, [Fe/H] and $v^{ini}_{rot}$, the
following quantities: MS lifetime ($\tau_{MS}$), mass of the He
core at the tip of the RGB phase (M$_{\rm H}^{\rm TIP}$),
He-burning lifetime ($\tau_{\rm He-burn}$), stellar age at the
beginning of the AGB phase ($t_{\rm AGB}$), mass of the CO core at
$t_{\rm AGB}$ (M$_{\rm CO}^{AGB}$), mass of the He core at
$\tau_{\rm AGB}$ (M$_{\rm H}^{AGB}$) and final central C/O ratio.
We note that the MS lifetime for the model with M=2 M$_\odot$,
[Fe/H]=0 and $v^{ini}_{rot}=120$ km/s is just 1\% longer than the
one of the corresponding non-rotating model, while in the low
metallicity set this difference becomes 2.5\%. This is a
consequence of the lifting caused by rotation during the MS. At
the RGB tip, when He-burning starts, the He-core mass of the model
with M=2.0 M$_\odot$, [Fe/H]=0 and $v_{rot}^{ini}$=120 km/s is
only 0.006 M$_\odot$ larger than that of the corresponding
non-rotating model. At low metallicity such a difference is 0.002
M$_\odot$. However, it is worth noting that the small variation of
the He-core mass at the RGB tip determines a slightly larger
luminosity during the Horizontal Branch phase. As a consequence,
the nuclear energy is released at a larger rate to balance the
higher surface radiative losses. Therefore, larger initial
rotation velocities imply higher central temperatures and, hence,
shorter He-burning lifetimes. In any case, the variation of
$\tau_{\rm He-burn}$ in our models is smaller than 5\%. As a
result, the central C/O ratio at the end of the core-He burning
phase is slightly larger in rotating models. At the beginning of
the AGB phase the physical and chemical structures of the rotating
models are slightly different from those of the corresponding
non-rotating models. In particular, increasing the initial
rotation velocity, the CO core mass grows up to $2\times 10^{-2}$
M$_\odot$ and $1.3\times 10^{-2}$ M$_\odot$ for the [Fe/H]=0 and
[Fe/H]=-1.7 sets, respectively. Similarly, the mass of the
H-exhausted core increases up to $1.5\times 10^{-2}$ M$_\odot$ and
$10^{-2}$ M$_\odot$, respectively.

\subsection{TP-AGB evolution and the $^{13}$C pocket: the [Fe/H]=0 case}

The changes caused by rotation on the evolution preceding the
TP-AGB phase are small enough that the development of thermal
pulses and the subsequent dredge up episodes are substantially
unaffected. Several rotation-induced instabilities affecting the
s-process nucleosynthesis, but not modifying the overall physical
properties of the models, arise during the TP-AGB phase.

\begin{figure}
\includegraphics[angle=0,width=\columnwidth]{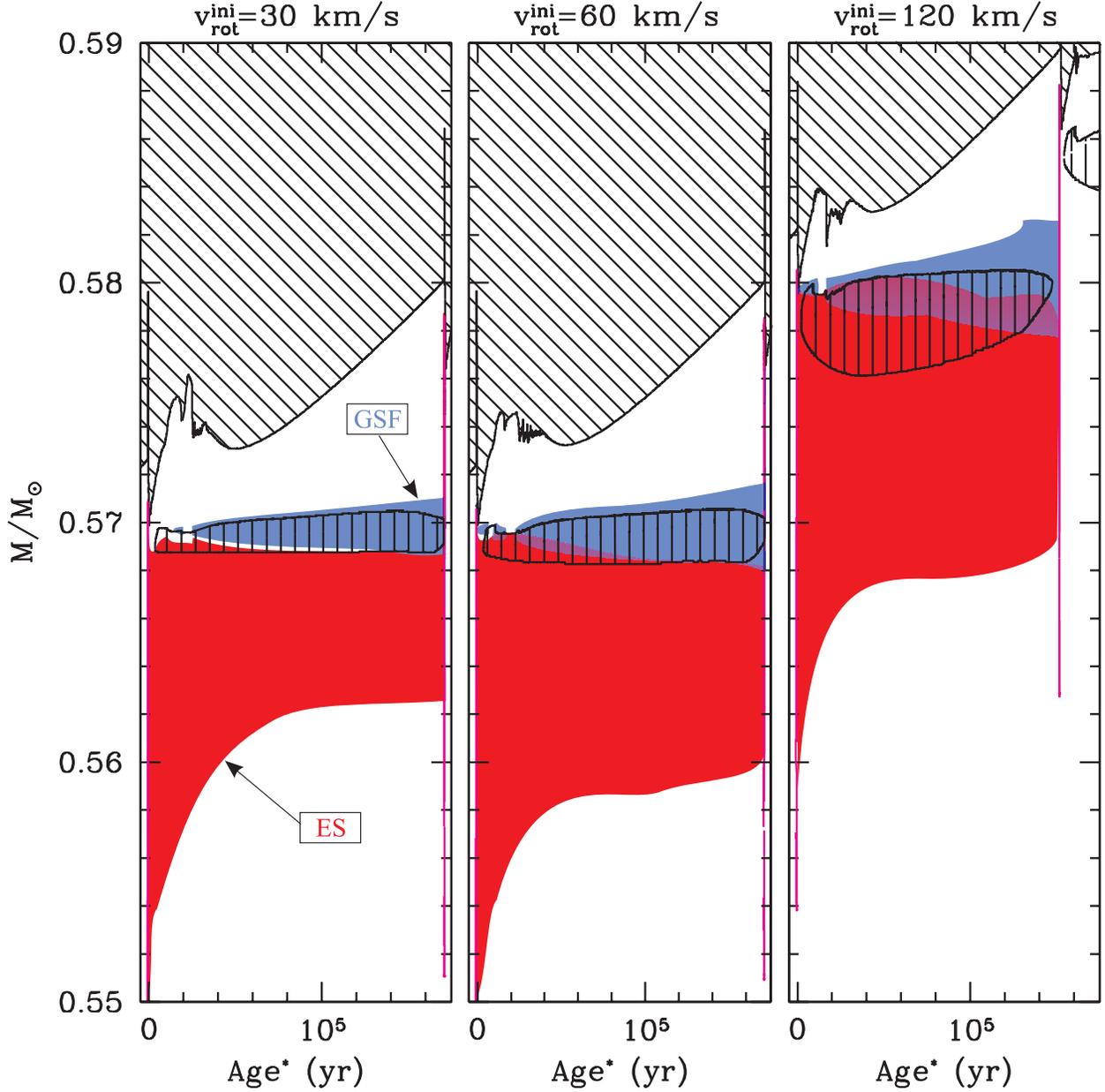}
\caption{Temporal evolution of rotation-induced instabilities
relevant for the s-process nucleosynthesis during the interpulse
period between the 2$^{nd}$ and the 3$^{rd}$ TPs of the 2
M$_\odot$ model with [Fe/H]=0 and different $v^{ini}_{rot}$. t=0
on the x-axis scale coincides with the epoch of the maximum
convective envelope penetration during the 2$^{nd}$ TDU episode.
Blue areas identify Goldreich-Schubert-Fricke unstable regions;
red areas identify Eddington-Sweet unstable regions. Magenta
spikes mark the convective shells powered by TPs. The oblique
shaded area highlights the convective envelope, while the vertical
shaded area is the region occupied by the
$^{13}$C-pocket.}\label{fig2}
\end{figure}

\begin{figure}
\includegraphics[angle=0,width=\columnwidth]{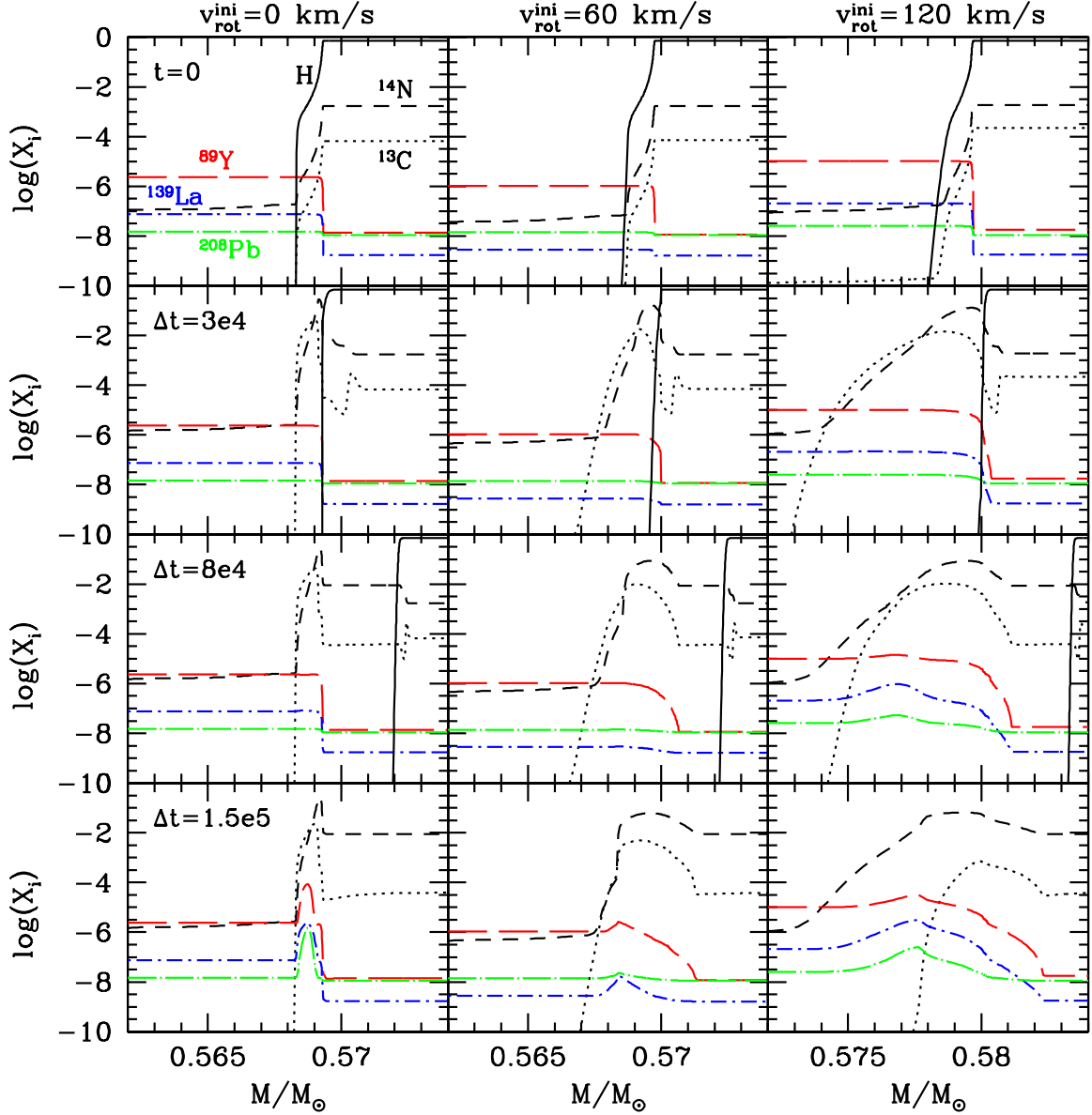}
\caption{Evolution of key selected isotopes in the region where
the $^{13}$C pocket forms during the interpulse after the 2$^{nd}$
TDU for the M=2 M$_\odot$, [Fe/H]=0 model with different initial
rotation velocities: H (solid), $^{13}$C (dotted), $^{14}$N
(short-dashed), $^{89}$Y (long-dashed), $^{139}$La
(dot-short-dashed) and $^{208}$Pb (dot-long-dashed) (see the
on-line edition for a color version).}\label{fig3}
\end{figure}

In order to describe the effects of angular momentum transport and
chemical mixing on the evolution of AGB models we illustrate the
formation and evolution of the 2$^{nd}$ $^{13}$C pocket, which
forms between the 2$^{nd}$ and the 3$^{rd}$ TPs, in 2 M$_\odot$
and [Fe/H]=0 models with different initial rotation velocities. In
Figure \ref{fig2} we report the evolution of the borders of the
rotation-induced unstable zones relevant for the s-process
nucleosynthesis as well as of convective regions during that
interpulse period. We also plot the $^{13}$C pocket (vertical
shaded region\footnote{We define this region as the layer where
X($^{13}$C)$>10^{-3}$.}). In Figure \ref{fig3} we plot the mass
fractions of selected key isotopes in the top layer of the
H-exhausted core, namely $^{1}$H (solid curve), $^{13}$C (dotted
curve), $^{14}$N (short-dashed curve), $^{89}$Y (long-dashed
curve), $^{139}$La (dot-short-dashed curve) and $^{208}$Pb
(dot-long-dashed curve). The four panels in each column refer to
the following epochs:
\begin{itemize}
{\item $t=0$, maximum penetration of the convective envelope
during the TDU;} {\item $t=3.0\times10^4$ yr, complete development
of the $^{13}$C pocket;} {\item $t=8.0\times 10^4$ yr, start of
the s-process nucleosynthesis;} {\item $t=1.5\times10^5$ yr, end
of the s-process nucleosynthesis.}
\end{itemize}

The occurrence of the TP drives the formation of a convective
shell which fully mixes the He-intershell region (magenta spikes
in Figure \ref{fig2}). When the TP quenches and this convective
shell disappears, the He-intershell is nearly homogeneous, so that
it becomes unstable to meridional circulations (red zone). Later
on, when the He-burning shell moves outward, the He-intershell
expands, the convective envelope (oblique shaded area) penetrates
inward and, then, recedes, leaving a variable H profile. The
resulting steep $\mu$-gradient acts as a barrier for meridional
circulations. In models with low $v^{ini}_{rot}$, the ES unstable
zone only marginally overlaps to the layer of variable hydrogen
profile. In any case, the circulation velocity is rather low.
Increasing the initial rotation velocity, $v_{ES}$ increases as
well as the extension of the overlapped layer, thus determining an
inward mixing of hydrogen (see first row in Figure \ref{fig3}).
Note that when the H mass fraction does not exceeds a few
$10^{-4}$, an incomplete CN cycle takes place, so that the
$^{13}$C production is favored with respect to the $^{14}$N
production. As a consequence, the inward mixing induced by
meridional circulations will imply more extended $^{13}$C pockets
(see second row in Figure \ref{fig3}). Since the meridional
circulations remain active during the whole interpulse, also the
products of the s process will be slowly mixed inward (see third
and forth rows in Figure \ref{fig3}).

\begin{figure}
\includegraphics[angle=0,width=\columnwidth]{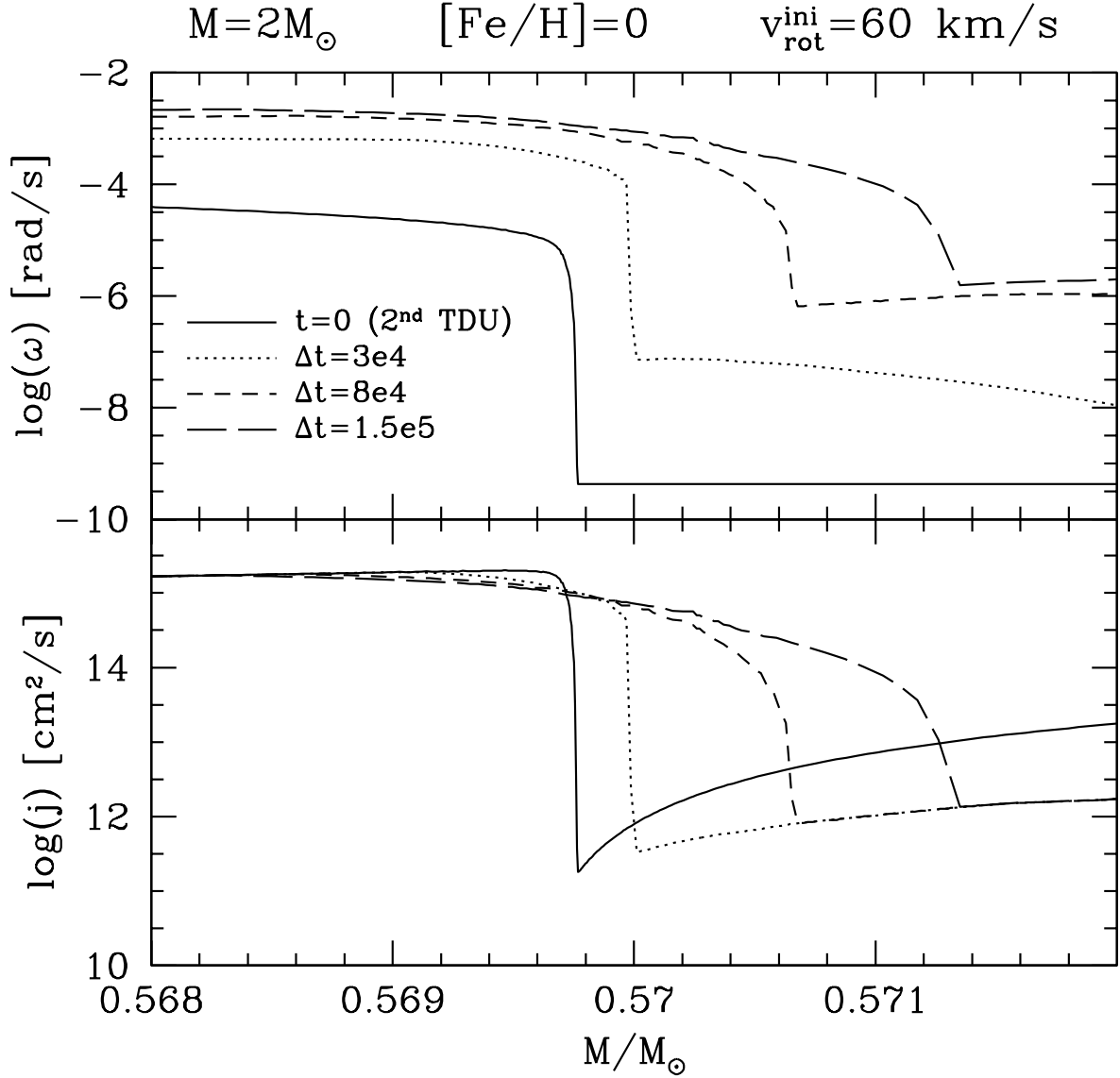}
\caption{Angular velocity (upper panel) and specific angular
momentum (lower panel) in the region interested by the 2$^{nd}$
TDU for the M=2 M$_\odot$, [Fe/H]=0 model with $v^{ini}_{rot}=60$
km/s. The various curves refer to different epochs (see text for
details).}\label{fig4}
\end{figure}

In addition to the meridional circulations another secular
instability takes place after the TDU (blue zone in Figure
\ref{fig2}). In Figure \ref{fig4} we report $\omega$ (upper panel)
and the specific angular momentum profiles (lower panel) for the
model with $v^{ini}_{rot}$=60 km/s. As shown in this Figure, a
sharp drop of the angular momentum develops in the transition zone
between the fast-rotating core and the slow-rotating envelope. The
drop of $j(m)$ triggers a GSF instability\footnote{This is the
same GSF instability already found and discussed in previous works
on rotating AGB models, ({\it e.g.} \citealt{he03,siess04}).}.
However, at the beginning of the interpulse period, the mixing and
the consequent redistribution of $j(m)$ are inhibited by the steep
gradient of mean molecular weight that separates the He-rich zone
from the H-rich envelope. As a consequence, the internal border of
the unstable zone remains almost fixed and close to the deepest
point previously attained by the fully convective envelope at the
time of the TDU. When the H burning restarts, a shallower $\mu$
gradient takes place and the GSF instability begins to move
inward. At that time, the $^{13}$C pocket is almost fully
developed. As shown in Figure \ref{fig2}, such a zone partially
overlaps with the $^{13}$C pocket. As the interpulse goes on, the
region affected by this instability grows in mass, due to the
outward diffusion of the angular momentum: $j(m)$ decreases at the
internal border, while it increases at the external one (see
Figure \ref{fig4}). For this reason the upper border of the
$^{13}$C pocket is progressively dragged upward. Since the GSF
instability is active in a $^{14}$N-rich zone, it produces a
contamination of the $^{13}$C pocket with a strong neutron poison
(see the second row in Figure \ref{fig3}). Also in this case, the
larger $v^{ini}_{rot}$ the higher $v_{GSF}$ and the more extended
the unstable zone (see Figure \ref{fig2}). Summarizing, in models
with $v^{ini}_{rot}\le60$ km/s, the s-process nucleosynthesis is
mainly affected by the GSF instability. On the contrary,
meridional circulations provide an important additional
modification of the $^{13}$C pocket in fast rotating models.

\begin{figure}
\includegraphics[angle=0,width=\columnwidth]{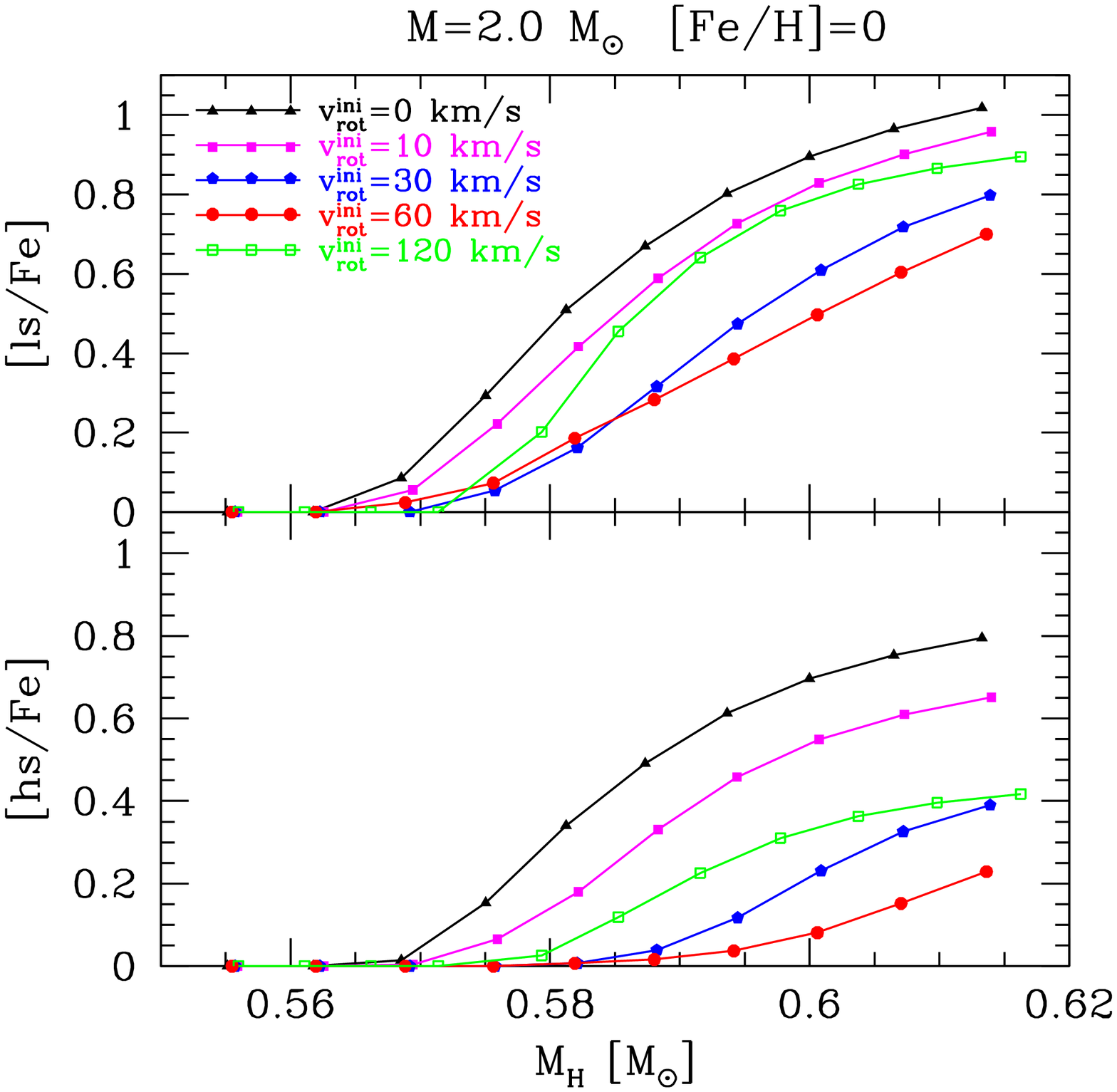}
\caption{Evolution of [ls/Fe] and [hs/Fe] as a function of the
core mass for the 2.0 M$_\odot$ model with [Fe/H]=0. Each curve
refers to a different initial rotation velocity (see the on-line
edition for a color version).}\label{fig5}
\end{figure}

\begin{figure}
\includegraphics[angle=0,width=\columnwidth]{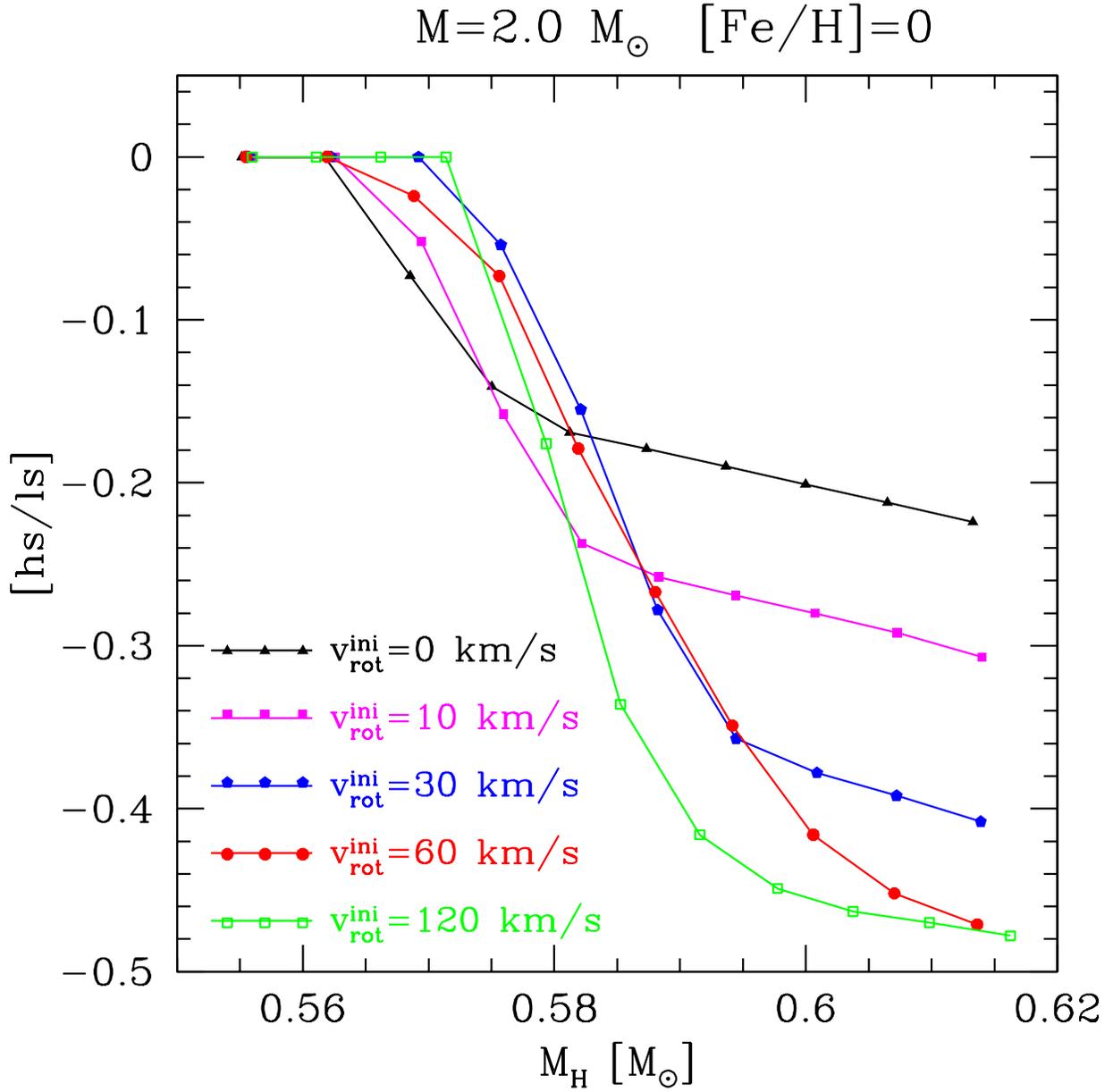}
\caption{The same as in Figure \ref{fig5}, but for the [hs/ls]
spectroscopic index (see the on-line edition for a color version
of this figure).}\label{fig6}
\end{figure}

The overall nucleosynthesis results of our calculations for the
solar metallicity set are displayed in Figures \ref{fig5} and
\ref{fig6}, where we report the evolution of
[ls/Fe]\footnote{[ls/Fe]=([Sr/Fe]+[Y/Fe]+[Zr/Fe])/3~.},
[hs/Fe]\footnote{[hs/Fe]=([Ba/Fe]+[La/Fe]+[Nd/Fe]+[Sm/Fe])/4~.}
and of the [hs/ls] index\footnote{[hs/ls]=[hs/Fe]-[ls/Fe]~.} at
the stellar surface. For $v^{ini}_{rot}\le 60$ km/s, the larger
the initial rotation velocity the smaller the final [ls/Fe] and
[hs/Fe]. This is a consequence of the $^{14}$N contamination of
the $^{13}$C pocket caused by the GSF instability. Moreover, the
hs component is more depleted with respect to the ls component.
The latter is a consequence of the lower neutrons-to-seeds ratio
determined by the combined actions of the two secular
rotation-induced instabilities previously described. Thus, the
[hs/ls] spectroscopic index decreases when increasing
$v^{ini}_{rot}$ (see Figure \ref{fig6}). The model with
$v^{ini}_{rot}$=120 km/s deviates from the described trend, as its
[ls/Fe] and [hs/Fe] result larger than those of the models with
$v^{ini}_{rot}$=30 km/s and $v^{ini}_{rot}$=60 km/s. Such an
occurrence is due to the particularly efficient meridional
circulations, which drives the formation of a more extended
$^{13}$C pocket. Note that, in any case, the [hs/ls] index
decreases as $v^{ini}_{rot}$ increases from 0 to 120 km/s.

\subsection{TP-AGB evolution and the $^{13}$C pocket: the [Fe/H]=-1.7 case}

\begin{figure}
\includegraphics[angle=0,width=\columnwidth]{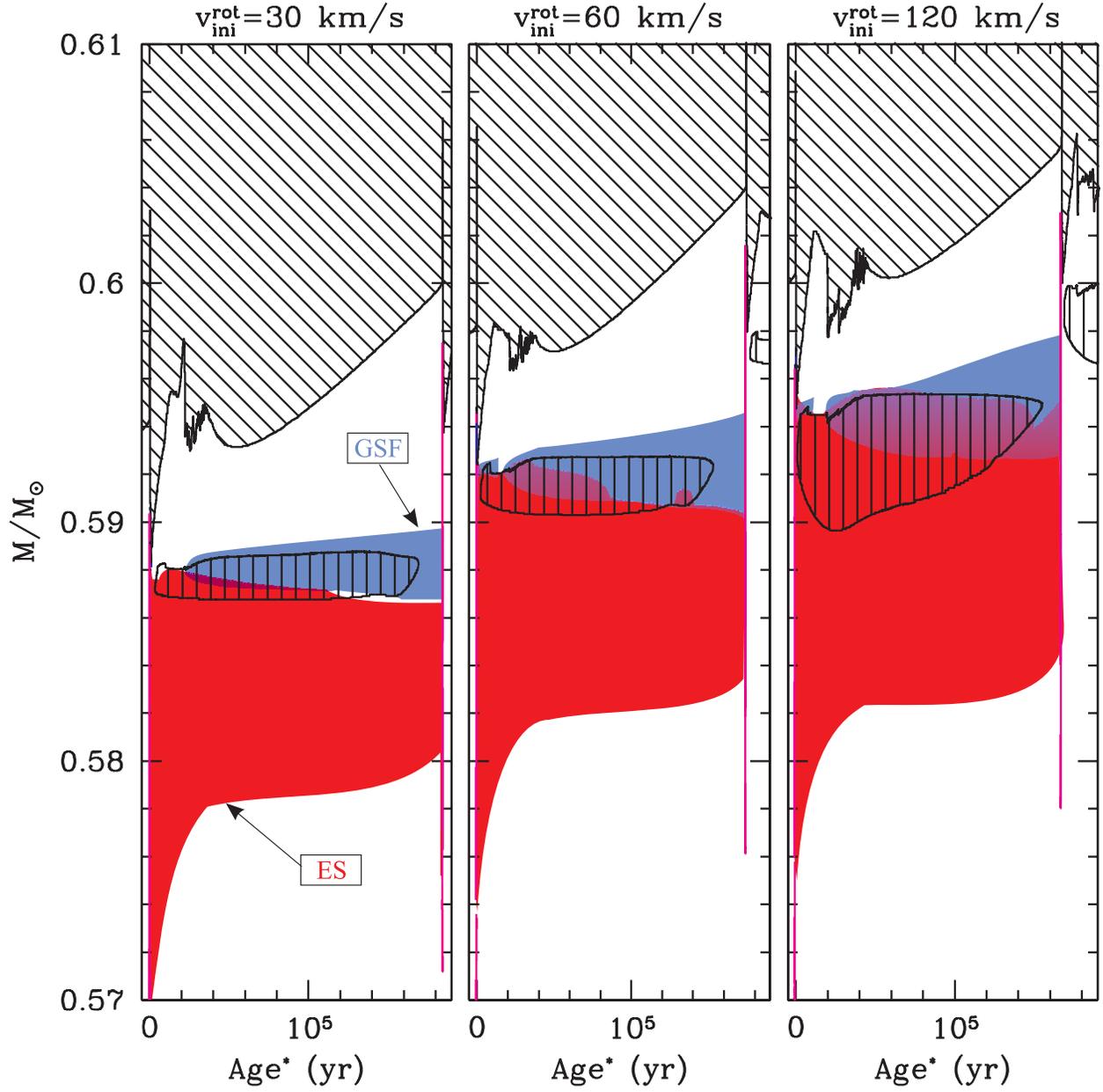}
\caption{The same as in Figure \ref{fig2}, but for the 1.5
M$_\odot$ model with [Fe/H]=-1.7.}\label{fig7}
\end{figure}

Low metallicity models exhibit the same global behavior, even if
some important differences there exist. In general, we expect, and
actually obtain as a result of our calculations, that the effects
of rotation are stronger at low metallicity because stellar
structures are more compact, so that for a given $v^{ini}_{rot}$
the corresponding angular velocity is larger. Moreover, since at
each TP some angular momentum is extracted from the He-intershell
and since the first efficient TDU episode occurs earlier at low
metallicity \citep[][and references therein]{cri09}, it comes out
that at the first TDU episode, the angular velocity in the region
where $^{13}$C pocket will form is larger. As a consequence, the
rotation-induced instabilities in low Z models are more efficient
and affect at a higher level the s-process nucleosynthesis. In
particular, the ES instability produces sizable effects even for
low $v^{ini}_{rot}$. This occurrence is shown in Figure
\ref{fig7}, where we report the same quantities as in Figure
\ref{fig2}, but for the 1.5 M$_\odot$  and [Fe/H]=-1.7 models.

\begin{figure}
\includegraphics[angle=0,width=\columnwidth]{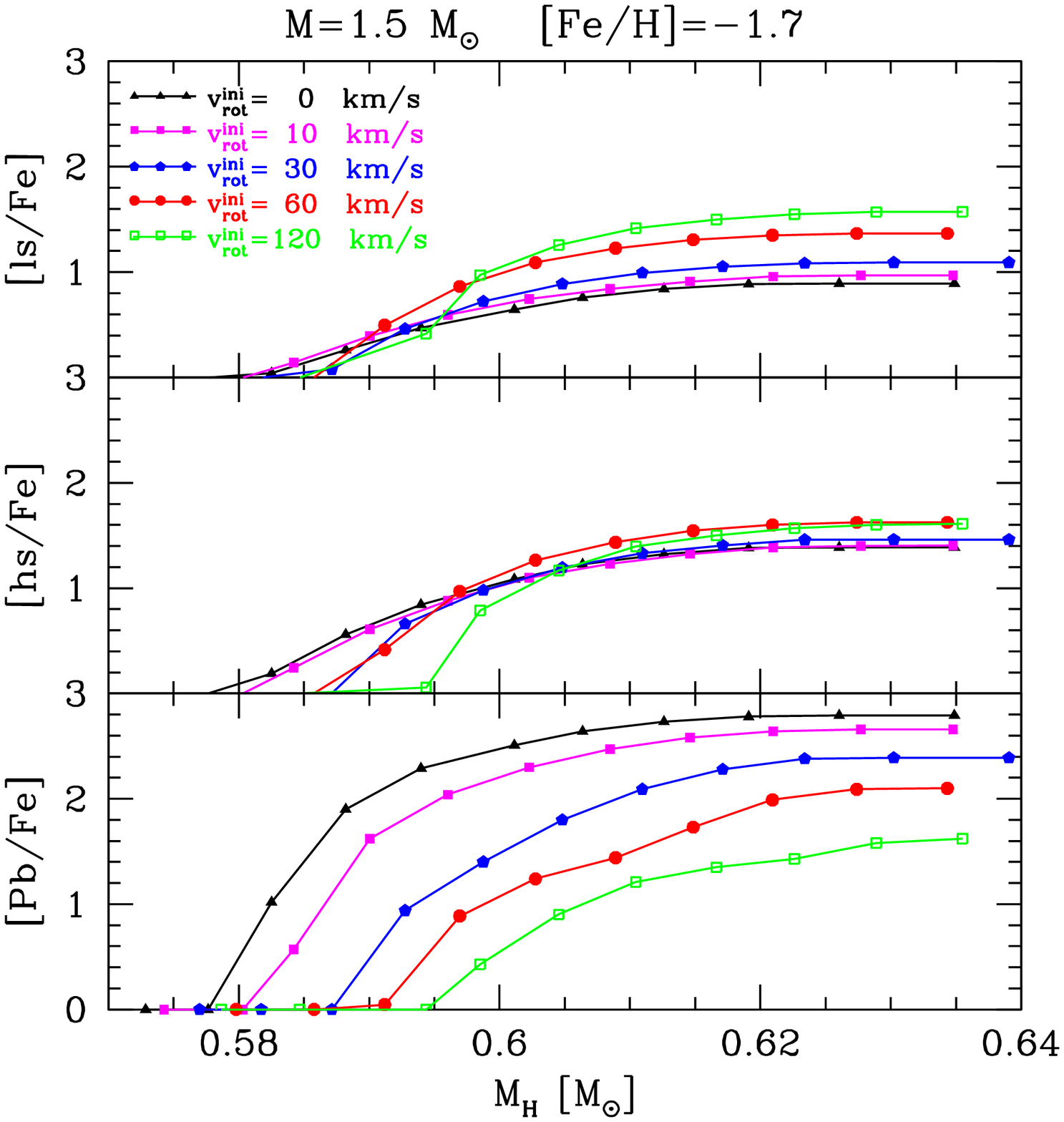}
\caption{Evolution of of [ls/Fe], [hs/Fe] and [Pb/Fe] as a
function of the core mass for the 1.5 M$_\odot$ model with
[Fe/H]=-1.7~. Each curve refers to a different initial rotation
velocity (see the on-line edition for a color
version).}\label{fig8}
\end{figure}

\begin{figure}
\includegraphics[angle=0,width=\columnwidth]{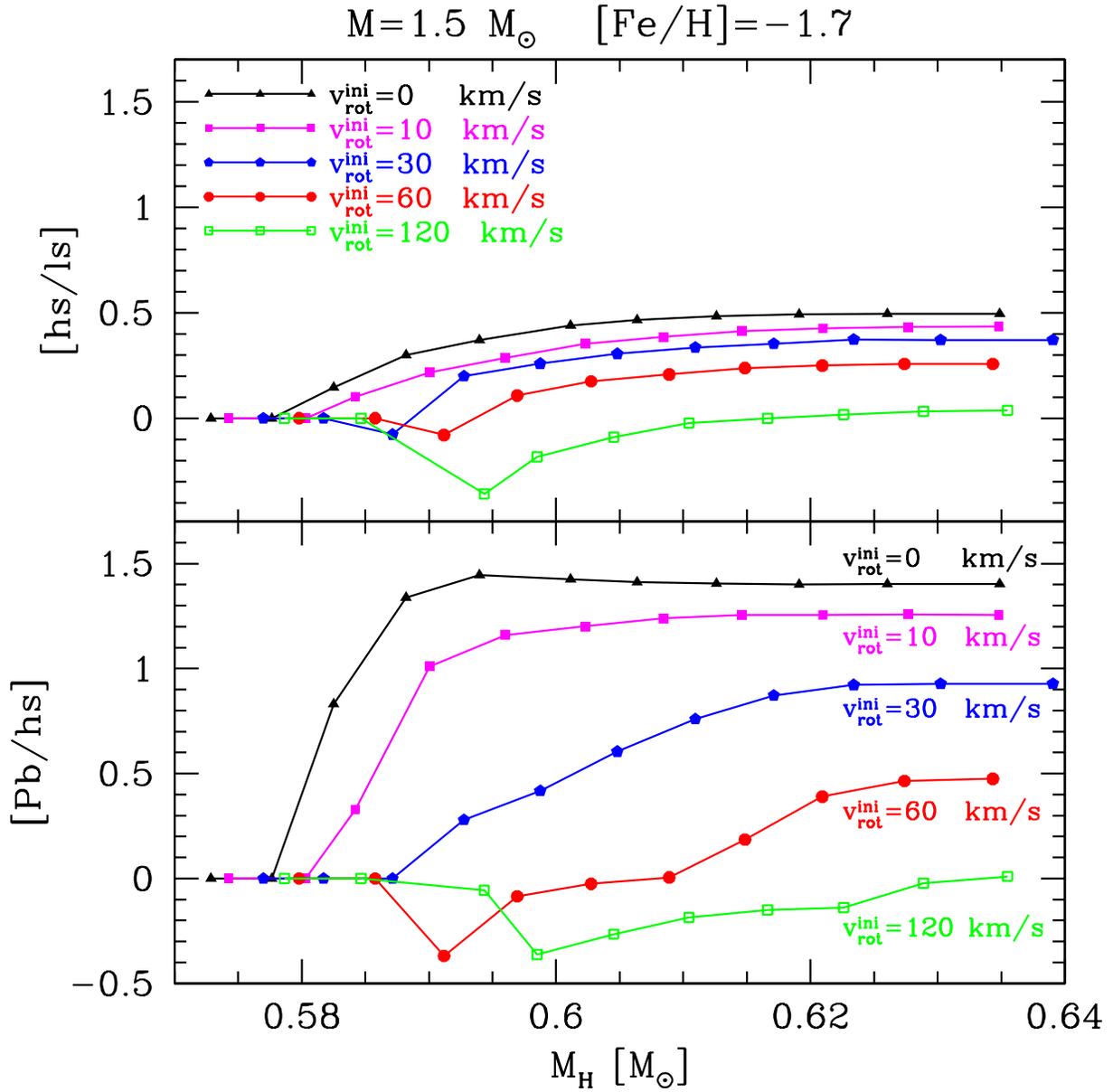}
\caption{The same as in Figure \ref{fig8}, but for the [hs/ls] and
[Pb/hs] spectroscopic indexes (see the on-line edition for a color
version).}\label{fig9}
\end{figure}

The nucleosynthesis results are illustrated in Figure \ref{fig8},
where we report the evolution of the [ls/Fe], [hs/Fe] and [Pb/Fe]
for all the computed models at [Fe/H]=-1.7. By increasing the
initial rotation velocity, the [Pb/Fe] decreases more than one
order of magnitude with respect to the non-rotating model, while
[ls/Fe] and [hs/Fe] increase by a factor of 2 and 5, respectively.
In low metallicities non-rotating models, the high
neutrons-to-seeds ratio favors the synthesis of a large amount of
lead with respect to the hs and, to a larger extent, the ls
components (see the discussion in \citealt{cri09}). Then, the
strong lead suppression and the corresponding increase of ls and
hs in rotating models are mainly due to the reduction of the
neutrons-to-seeds ratio caused by both the GSF and the ES
instabilities. For the same reason, the [hs/ls] decreases by a
factor of $\sim$3 while the
[Pb/hs]\footnote{[Pb/hs]=[Pb/Fe]-[hs/Fe]~.} decreases of more than
a factor 20, reaching a value [Pb/hs]$\sim$0 in the
model with $v^{ini}_{rot}$=120 km/s (see Figure \ref{fig9}).\\

Note that, as for the solar Z models, the total amount of heavy
elements is, in any case, smaller in rotating models. This is a
consequence of the reduction of the total amount of neutrons
available for the s process, as due to the $^{14}$N contamination
of the $^{13}$C pocket induced by the GSF instability.

\section{Uncertainties on the angular momentum transport and on the treatment of rotation-induced mixing}
\label{unce}

\begin{figure}
\includegraphics[angle=0,width=\columnwidth]{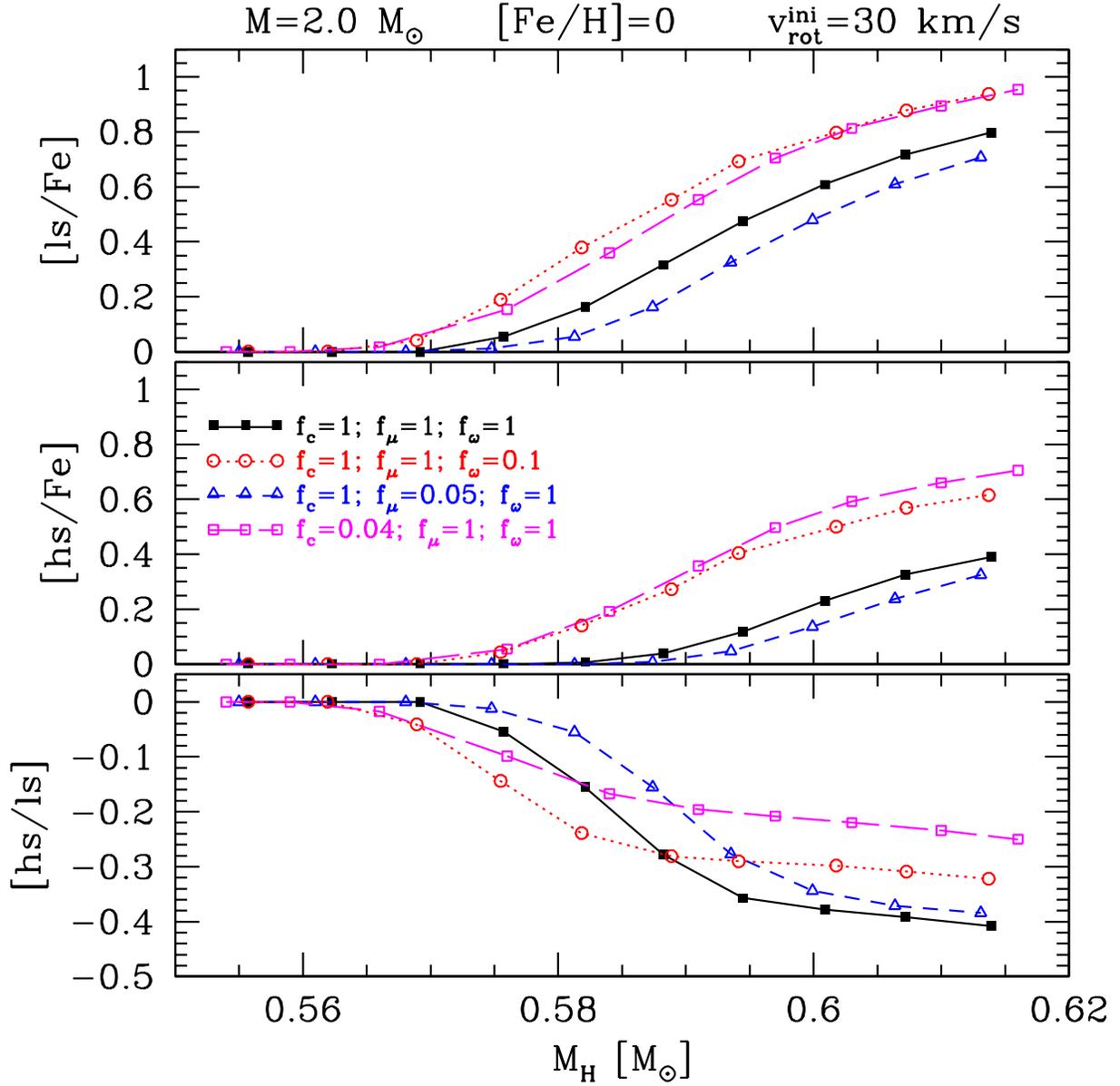}
\caption{Evolution of [ls/Fe], [hs/Fe]  and [hs/ls] for the 2.0
M$_\odot$ model with [Fe/H]=0 and $v^{ini}_{rot}$=30 km/s, with
different choices for the free parameters (see the on-line edition
for a color version).}\label{fig10}
\end{figure}

\begin{figure}
\includegraphics[angle=0,width=\columnwidth]{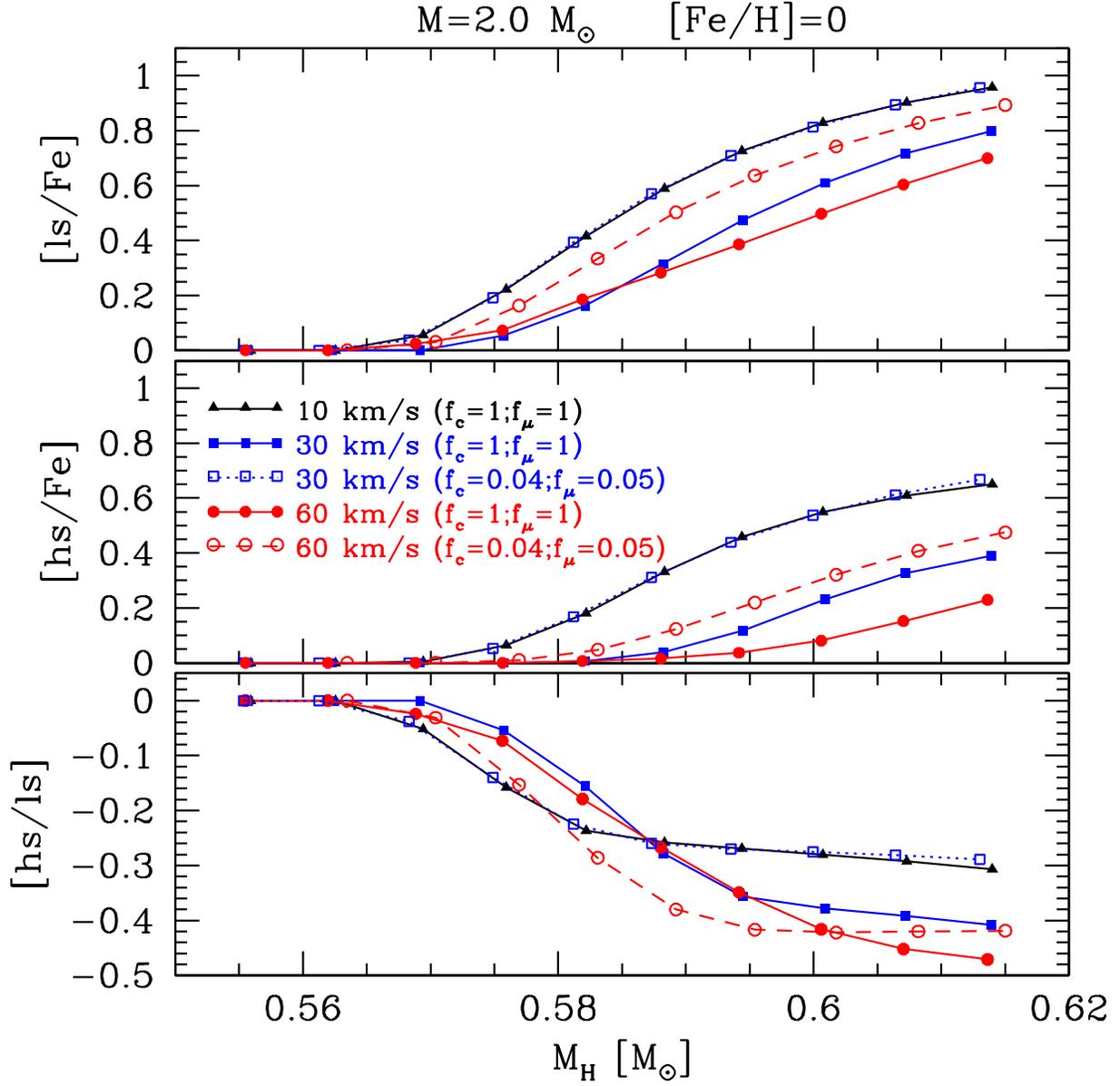}
\caption{The same as in Figure \ref{fig11}, but for different
initial rotation velocities and free parameters combinations (see
the on-line edition for a color version).}\label{fig11}
\end{figure}

As already discussed in Section \ref{algo}, the treatments of both
the angular momentum transport and the rotation-induced mixing are
affected by many uncertainties. In order to evaluate their effects
on the final heavy elements s process surface distributions, we
compute an additional set of models with M=2 M$_\odot$, [Fe/H]=0
and $v^{ini}_{rot}$=30 km/s, by adopting different values for the
$f_\mu$, $f_c$ and $f_\omega$ parameters introduced in Equations
(\ref{eqdj}), (\ref{eqdc}) and (\ref{eqvmu}). Our result are
summarized in Figure \ref{fig10}, where we report the [ls/Fe],
[hs/Fe] and [hs/ls]. As a result, a general reduction of the
efficiency of both the angular momentum transport and mixing of
chemical species (model with $f_c$=$f_\mu$=1 and $f_\omega$=0.1 :
open circles with dotted lines) reduces the effect of rotation so
that both the ls and hs elements are overproduced with respect to
the reference case ($f_c$=$f_\mu$=$f_\omega$=1 : filled squares
with solid lines). By recalling that at solar metallicity, for
moderate values of the initial rotation velocity, the net effect
of rotation-induced instabilities is a larger depletion of the hs
component, it comes out that the model with $f_\omega$=0.1 attains
a slightly higher [hs/ls] (lower panel of Figure \ref{fig10}). The
same conclusion is still valid for the model with
$f_\mu$=$f_\omega$=1 and $f_c$=0.04 (open squares with long-dashed
lines). Since the $f_c$ parameter affects only the mixing of
chemical species but not the transport of angular momentum, by
comparing this model to the one with $f_c$=$f_\mu$=1 and
$f_\omega$=0.1 we conclude that the s-process nucleosynthesis
during the AGB phase is not significantly affected by the
uncertainty related to the angular momentum transport. In this
regard, it is important to note that for the model with $f_c$=0.04
the mixing efficiency of chemical species has been reduced by a
factor larger than in the model with $f_\omega$=0.1~. As a
consequence, the corresponding [hs/Fe] results slightly larger,
thus determining an increase of the [hs/ls] index.\\
In order to test the sensitivity of the s process to the
inhibiting effect of $\mu$ gradients on secular instabilities (ES
and GSF), we compute a model with $f_c$=$f_\omega$=1 and
$f_\mu$=0.05. Owing to the reduction of the $\mu$-gradient
barrier, the zones interested by both the GSF and the ES
instabilities are larger and the effective circulation velocities
increase. As a result, the final amount of ls and hs elements is
lower than what we find in the $f_\mu=1$ case, but the final
[hs/ls] is only marginally modified (open squares with long dashed
lines). It is important to remark that a decrease of $f_\mu$ is
not equivalent to an increase of the efficiency of both angular
momentum transport and mixing of chemical species ({\it e.g.} as
obtained by increasing $f_\omega$). In fact, on a general ground,
the reduction of $\mu$-currents determines not only the increase
of the efficiency of rotation-induced instabilities, but
also the enlargement of the unstable zones.\\
Finally, we have computed two additional models by fixing
$f_\omega=1$, $f_\mu=0.05$ and $f_c=0.04$ for two different values
of the initial rotation velocity, namely $v^{ini}_{rot}=30$ km/s
and $v^{ini}_{rot}=60$ km/s. The obtained results are displayed in
Figure \ref{fig11}, where we report the evolution of the surface
[ls/Fe], [hs/Fe] and [hs/ls]. For comparison we also plot the same
quantities for models with $f_c=f_\mu=f_\omega=1$ and
$v^{ini}_{rot}=10$ km/s (filled triangles with solid line),
$v^{ini}_{rot}=30$ km/s (filled squares with solid line) and
$v^{ini}_{rot}=60$ km/s (filled circles with solid line). By an
inspection to Figure \ref{fig11} it comes out that models with
reduced $f_c$ and $f_\mu$ mimic models with lower initial rotation
velocities and $f_c=f_\mu=1$. As a matter of fact, the overall
uncertainties produced by varying the three free parameters do not
exceed the variations of the surface chemical abundances obtained
by changing the initial rotation velocity within the range
explored in this paper.

Finally, we have computed a few additional models in order to
reproduce the mean-core-rotation period (20-200 days) derived by
\citet{mosser12} for a sample of red clump stars with solar-like
metallicity. We found that the efficiency of angular momentum
transport should be increased by a factor of at least 1000. In
practice, we are able to reproduce the asteroseismic data by
setting $f_\omega> 1000$ and, accordingly, by reducing $f_c$ in
order to maintain $f_\omega\cdot f_c=1$. In this way, with respect
to our standard model ($f_\omega=f_c=1$), the mean angular
velocity in the He-core decreases by roughly a factor 200 down to
1.4$\times 10^{-6}$ rad/s. In such a model, at the onset of the
TP-AGB phase the angular velocity in the He-intershell is also two
orders of magnitude smaller than in the standard case. Thus, the
ES and GSF induced mixing marginally modifies the mass extensions
of the $^{13}$C and $^{14}$N pockets and their relative overlap.
As a consequence, the resulting s-process surface enrichment is
very similar to the non-rotating model. We will further comment
this result in the next Section.

\section{Summary and Conclusions}

We investigated the effects of rotation on the s-process
nucleosynthesis in low-mass AGB stars by computing full
evolutionary models from the pre-Main Sequence up to the AGB tip.
In our computations we included the angular momentum transport and
the mixing of chemical species, as determined by convection and
rotation-induced instabilities. We also discussed the
uncertainties affecting the efficiencies of angular momentum
transport and mixing.

In Table \ref{tab2}, we report the final values of [ls/Fe],
[hs/Fe], [Pb/Fe], [hs/ls] and [Pb/hs] for all the computed models.
The evolution of the abundances of elements and isotopes as well
as the chemical yields related to the models presented in this
paper will be available on the FRUITY database\footnote{Available
at {\it fruity.oa-teramo.inaf.it}.} \citep{cris11}. In the same
Table we also report the final carbon surface enhancements for the
computed models, assumed as representative of the mass globally
dredged up during the AGB phase. It is worth mentioning that the
inclusion of rotation does not substantially alter the surface
[C/Fe], thus confirming that its effects on the TDU efficiency are
minimal. The largest differences characterizing models with
$v^{ini}_{rot}$=120 km/s can be safely attributed to the larger
core masses at the beginning of the AGB phase (see
\citealt{stra03b} for a detailed description of the relation
between core mass and TDU efficiency).

Our analysis shows that a variation in the initial velocity
determines a consistent spread in the final surface s-process
enhancements and spectroscopic indexes in stars with the same
initial mass and metallicity. Rotation-induced instabilities
modify the mass extension of both the $^{13}$C and the $^{14}$N
pockets and their overlap. The average neutrons-to-seeds ratio is
also reduced. This has two major consequences: i) the total amount
of heavy elements produced by the s process is lower and ii) the
surface abundances are redistributed in such a way to favor the
light-s elements with respect to the heavier ones. As a matter of
fact, both the [hs/ls] and the [Pb/hs] spectroscopic indexes
decrease as the initial rotation velocity increases. Our results
suggest that rotation can be regarded as a possible physical
mechanism responsible for the observed spread of the [hs/ls]
index. Particularly interesting is the strong reduction of [Pb/hs]
in rotating models. In this regard, we recall that the origin of
lead in the solar system is usually ascribed to low metallicity
AGB stars. Theoretical models show that the maximum lead
production occurs at Z$\sim 10^{-3}$. Rare are lead measurements
in low metallicity AGB and post-AGB stars belonging to the
Galactic disk. On the other hand, a direct observation of Halo AGB
stars presently undergoing TDU is not possible as they have
already evolved to their cooling sequence. However, the imprint of
this ancient stellar population can be detected in
Carbon-Enhanced-Metal-Poor stars enriched in s elements. Most of
these stars are heavily enriched in Pb, even if some of them show
a relatively low [Pb/hs] (see \citealt{bi11}). Also in this case
rotation can be regarded as a possible explanation of the observed
spread in the heavy elements distributions.

The obtained results depend on the efficiency of angular momentum
transport during the pre-AGB evolution. In particular, if the
inner zones of a rotating star are efficiently spun down, as
suggested by a recent analysis of rotational splittings in a
sample of RGB and red clump stars with solar-like metallicity
observed by Kepler, the rotation induced instabilities during the
TP-AGB phase do not significantly alter the s-process
nucleosynthesis. Note, however, that these observations are still
rather limited in mass and chemical composition. In particular, we
do not know if such a spin down of the core is so efficient also
at the metallicity of Halo stars.\\
Further observational constraints on the efficiency of angular
momentum transport in low- and intermediate-mass stars can be
derived from the rotational broadening of the Ca II K line in WDs
\citep{berger05}. Rotation velocities up to 10 km/s, with an
average value of about 1 km/s, have been observed. By adopting the
same procedure described by \citet{suijs08} to derive the WDs
rotation velocity from the last computed model at the tip of the
AGB, we find that our standard model with M=2 M$_\odot$, [Fe/H]=0
and $v^{ini}_{rot}$=30 km/s produces a WD of M= 0.63 M$_\odot$
with $v^{WD}_{rot}=58$ km/s. The same model, but with
$f_\omega>1000$ (see \S \ref{unce}), leads to $v^{WD}_{rot}<0.3$
km/s. Therefore, if models preserving the core angular momentum
predict too short WDs rotation periods, those in which the core is
substantially spun down, as required by asteroseismic observations
of red clump stars, predict rather long WDs rotation periods.

\acknowledgments

We thank the anonymous referee for a careful and constructive
reading of the manuscript. We acknowledge support from the Italian
Ministry of Education, University and Research under the FIRB2008
program (RBFR08549F-002) and from the PRIN-INAF 2011 project
"Multiple populations in Globular Clusters: their role in the
Galaxy assembly".

\bibliographystyle{aa}
\bibliography{Piersanti_AGB_rot}

\begin{deluxetable}{cccccccccccc}\rotate
\tablecolumns{12} \tablewidth{0pc} \tablecaption{Physical
properties of the computed models (see text for details).}
\tablehead{\colhead{[Fe/H]} & \colhead{M} &
\colhead{$v^{ini}_{rot}$} & \colhead{$\omega_0$} &
\colhead{$J_0$}& \colhead{$\tau_{\rm MS}$} & \colhead{M$_{\rm
H}^{\rm TIP}$} & \colhead{$\tau_{\rm He-burn}$} & \colhead{$t_{\rm
AGB}$} & \colhead{M$_{\rm CO}^{\rm AGB}$} &
\colhead{M$_{\rm H}^{\rm AGB}$} & \colhead{C/O}\\
 & \colhead{[M$_\odot$]} &
\colhead{[km/s]} & \colhead{[$10^{-4}$ $rad/s$]} &
\colhead{[$10^{50}$ g $cm^2/s$]} & \colhead{[$10^8$ yr]}
&\colhead{[M$_\odot$]} &\colhead{[$10^8$ yr]} &\colhead{[$10^8$
yr]} &\colhead{[M$_\odot$]} &\colhead{[M$_\odot$]} &}\startdata
0    & 2.0  &   0 &  0.00  & 0.00 & 9.693 & 0.4628  &  1.212 & 12.240 & 0.512 & 0.545 & 0.253\\
0    & 2.0  &  10 &  0.09  & 0.22 & 9.702 & 0.4651  &  1.196 & 12.248 & 0.513 & 0.545 & 0.255\\
0    & 2.0  &  30 &  0.28  & 0.65 & 9.706 & 0.4653  &  1.194 & 12.256 & 0.517 & 0.550 & 0.257\\
0    & 2.0  &  60 &  0.56  & 1.30 & 9.714 & 0.4659  &  1.190 & 12.279 & 0.519 & 0.550 & 0.257\\
0    & 2.0  & 120 &  1.09  & 2.61 & 9.783 & 0.4686  &  1.147 & 12.385 & 0.532 & 0.561 & 0.267\\
-1.7 & 1.5  &   0 &  0.00  & 0.00 & 16.470 & 0.4794 &  0.913 & 19.973 & 0.548 & 0.579 & 0.394\\
-1.7 & 1.5  &  10 &  0.14  & 0.14 & 16.468 & 0.4795 &  0.913 & 19.970 & 0.545 & 0.574 & 0.395\\
-1.7 & 1.5  &  30 &  0.42  & 0.43 & 16.493 & 0.4797 &  0.908 & 19.984 & 0.548 & 0.577 & 0.399\\
-1.7 & 1.5  &  60 &  0.84  & 0.86 & 16.542 & 0.4799 &  0.900 & 20.050 & 0.552 & 0.581 & 0.404\\
-1.7 & 1.5  & 120 &  1.70  & 1.70 & 16.905 & 0.4811 &  0.885 & 20.474 & 0.561 & 0.587 & 0.406\\
\enddata \label{tab1}
\end{deluxetable}

\begin{deluxetable}{ccccccccc}
\tablecolumns{9} \tablewidth{0pc} \tablecaption{Surface s-process
enhancements, s-process indexes and [C/Fe] for all the computed
models.} \tablehead{\colhead{[Fe/H]} & \colhead{M (M$_\odot$)} &
\colhead{$v^{ini}_{rot}$ ($km/s$)} & \colhead{[ls/Fe]} &
\colhead{[hs/Fe]} & \colhead{[Pb/Fe]} & \colhead{[hs/ls]} &
\colhead{[Pb/hs]} & \colhead{[C/Fe]}} \startdata
0    & 2.0  &   0 & 1.02 &  0.80  & 0.50  & -0.22   &  -0.30  &  0.52  \\
0    & 2.0  &  10 & 0.96 &  0.65  & 0.35  & -0.31   &  -0.30  &  0.51  \\
0    & 2.0  &  30 & 0.80 &  0.39  & 0.13  & -0.41   &  -0.26  &  0.51  \\
0    & 2.0  &  60 & 0.70 &  0.23  & 0.05  & -0.47   &  -0.18  &  0.51  \\
0    & 2.0  & 120 & 0.90 &  0.42  & 0.06  & -0.48   &  -0.36  &  0.46  \\
-1.7 & 1.5  &   0 & 0.89 &  1.39  & 2.79  &  0.50   &  1.40   &  2.31  \\
-1.7 & 1.5  &  10 & 0.97 &  1.41  & 2.66  &  0.44   &  1.26   &  2.28  \\
-1.7 & 1.5  &  30 & 1.09 &  1.46  & 2.39  &  0.37   &  0.93   &  2.28  \\
-1.7 & 1.5  &  60 & 1.37 &  1.63  & 2.09  &  0.26   &  0.47   &  2.26  \\
-1.7 & 1.5  & 120 & 1.57 &  1.62  & 1.62  &  0.04   &  0.01   &  2.20  \\
\enddata
\label{tab2}
\end{deluxetable}

\end{document}